\documentclass[preprint,aps]{revtex4}
\usepackage{graphicx}
\usepackage{float}
\begin{document}
\title{Determination of the single particle distribution function in a weakly correlated weakly inhomogeneous plasma}
\author{Anirban Bose}
\affiliation{ Serampore College, Serampore, Hooghly 712201, India}

\begin{abstract}
Single particle distribution function of plasma particles has been derived from the first member of the Bogoliubov-Born-Green-Kirkwood-Yvon (BBGKY) hierarchy utilising the pair correlation function evaluated in \cite{kn:ab1} from the second member of the BBGKY hierarchy. This distribution function may be employed to probe the thermodynamic properties of the weakly inhomogeneous plasma systems.
\end{abstract}

\maketitle
Pair correlation function plays a significant role in studying the thermodynamic properties of plasma. Therefore, the determination and application of pair correlation function in finding out the properties of plasma has been the topic of interest for a long time. The method proposed by Dupree \cite{kn:dup} in case of slowly time and space varying system has been sucessfully applied to demonstrate the pair correlation function in a dilute, uniform, quiescent plasma \cite{kn:wolf}. By using path integral technique the quantum pair correlation function has been calculated in a one component plasma \cite{kn:jan}. The thermodynamic property of strongly coupled classical plasma has been investigated by nodal expansion \cite{kn:fur}.  interaction potential and thermodynamic functions of dusty plasma has been investigated  by measured correlation functions \cite{kn:fortov}. Study of pair correlation function using molecular dynamics simulation of the strongly coupled two temperature plasma is reported \cite{kn:nrs}.
parametrization of the pair correlation function and the static structure factor of the Coulomb one component plasma(OCP) is presented from the weakly coupled regime to the strongly coupled regime \cite{kn:des}.  The pair distribution function of both strongly coupled and strongly degenerate Fermi systems have been calculated using a novel path integral representation \cite{kn:fil} of the many-particle density operator. Recently, Monte Carlo simulations have been utilized to study distribution functions of the warm dense uniform electron gas in the thermodynamic limit \cite{kn:do}.

The statistical mechanics of plasma by using the Bogoliubov-Born-Green-Kirkwood-Yvon chain of equations has been attempted by many researchers in the past. In this connection O'neil and Rostoker \cite{kn:neil} have considered the plasma system in thermal equilibrium and evaluated the two and three body correlation function in homogeneous plasma. Using \cite{kn:da} the BBGKY hierarchy, the quantum binary and
triplet distribution functions for a neutral many-component
plasma has been determined. Previously, in a series of papers \cite{kn:ab,kn:ab1} an equation of pair correlation function was established from the first two members of BBGKY hierarchy in the weakly correlated limit and subsequently an expression of pair correlation function was derived in the weakly inhomogeneous limit. In this article we shall discuss the effect of the pair correlation function on the single particle distribution function. In order to do this we should insert the pair correlation function in the first member of BBGKY hierarchy and try to identify the contribution as some measureable macroscopic quantity to the potential energy of the system. 

We have considered a plasma system of N electrons and N infinitely massive ions which are randomly distributed in a volume V. The plasma is in thermal equilibrium.  

In this article, the system is considered to be weakly inhomogeneous. The inhomogeneous density of electrons is considered as
\begin{eqnarray}
n(3)= n_{0}+n_{1}(3) 
\end{eqnarray}
$n_{0}$ is the homogeneous part and the inhomogeneous part is $n_{1}(3)=B\cos (\textbf{p}\cdot \textbf{x}_{3})$.

The average interaction potential between two electrons separated
by a distance
 $r= \mid {\bf {x_1-x_2}} \mid $ is assumed to be of the Debye-H\"{u}ckel type\cite{kn:akh}:
\begin{equation}\phi_{12}({\bf {x_1-x_2}})= \frac{e^2}{r}{\rm exp}(-k_Dr)\label{v1}\end{equation}
The electronic charge is e and $T$ is the absolute temperature of the electrons.
  $$k_D^2 =n e^2/k_B T, $$
As we are interested in the inhomogeneous case, the density $n$ is space dependent and for a pair of particles with the space coordinates $\textbf{x}_{1}$ and $\textbf{x}_{2}$ we may choose the density $n$ in the expression of $k_{D}$ at the mid position $\textbf{r}_{m}$ ($\textbf{r}_{m}=\frac{\textbf{x}_{i}+\textbf{x}_{j}}{2}$) of the concerned particles in the first approximation.

Hence,  
\begin{equation} n(\textbf{r}_{m})=n_{0}+B\cos[\frac{\mathbf{p}}{2}\cdot(\mathbf{x}_{1}+\mathbf{x}_{2})]\end{equation}
In the weakly inhomogeneous case ($B\ll n_{0}$), $\phi_{12}$ as given by eq.(\ref{v1}) may be expressed as
\begin{equation}\phi_{12}= \frac{e^2}{r}{\rm exp}(-k_{D0}r)(1-B\frac{k_{D0}r}{2n_{0}}\cos[\frac{\mathbf{p}}{2}\cdot(\mathbf{x}_{1}+\mathbf{x}_{2})])\label{v2}\end{equation}
In the above expression we have neglected terms containing $(\frac{B}{n_{0}})^{s}$ when $s\geq 2$.

The first member of BBGKY hierarchy is given by \cite{kn:akh}
\begin{equation} \frac{\partial f_1}{\partial t} + {\bf v_1}\cdot\frac{\partial f_1}{\partial {\bf x_1}}
+ n_0\int d{\bf X_2} {\bf a_{12}}\cdot\frac{\partial}{\partial{\bf
v_1}}\left [ f_1({\bf X_1})f_1({\bf X_2})+g_{12}\right ] = 0
\end{equation}
For the electrons in equilibrium ($\frac{\partial f_1}{\partial t}=0$), the first member of BBGKY hierarchy is
\begin{equation}
{\bf v_{1}} \cdot \frac{\partial f_{1}}{\partial {\bf
x_{1}}}+n_{0}\int d{\bf X_2} f_{1}({\bf X_2}){\bf a_{12}}\cdot
\frac{\partial f_{1}({\bf X_1})}{\partial {\bf v_{1}}}+n_{0}\int
 d{\bf X_2}{\bf a_{12}}\cdot \frac{\partial g_{12}}{\partial
{\bf v_{1}}}=0\label{v3}\end{equation} where ${\bf a_{12}}
=-(1/m){\partial\phi_{12}}/{\partial {\bf x_1}} $ with $\phi_{12}$
given by eq.(\ref{v2}). 
The single
particle distribution functions are written in the follwing form:
$$f_1 ({\bf X_1}) = f_M({\bf v_1})F_1({\bf x_1})  $$
where $f_M$ is a Maxwellian distribution and $F_1$ denotes the
spatially dependent part of the single particle distribution function. 
 For a  plasma in  equilibrium, the pair correlation
function $g_{12}$ is written in  the form
\begin{eqnarray}
 g_{12}({\bf {X_1,X_2}})=f_{1}({\bf X_1})f_{1}({\bf X_2})\chi_{12}({\bf x_1,x_2})
\end{eqnarray}
$\chi_{12}({\bf x_1,x_2})$ is a symmetric function of $\bf x_1$ and $\bf x_2$. 
The expression of $\chi_{12}$ is obtained in \cite{kn:ab1} and it can be expressed as
\begin{equation}\chi_{12}= -\frac{\phi_{12}}{k_B T}\end{equation}
where $\phi_{12}$ is given by eq.(\ref{v2}).

The last term of eq.(\ref{v3}) is given
by

\begin{eqnarray} && n_{0}\int
 d{\bf X_2}{\bf a_{12}}\cdot \frac{\partial g_{12}}{\partial
{\bf v_{1}}} = n_0 \int d{\bf X_{2}} \left (
-\frac{1}{m}\frac{\partial}{\partial {\bf x_1}}{\phi_{12}}\right )
\cdot \frac{\partial} {\partial{\bf v_{1}}}\left (
-\frac{\phi_{12}}{k_B T}f_{1}({\bf X_1})f_{1}({\bf X_2}) \right )
\nonumber \\
&=& \frac{n_0}{mk_B T}  \int  d{\bf X_2}\nabla_{\bf x_1}
\phi_{12}^2 \cdot \frac{\partial f_{1}({\bf X_1})}{\partial {\bf
v_1}}f_{1}({\bf X_2})
\nonumber \\
&=& \frac{n_0}{mk_B T}  \int  d{\bf X_2}\nabla_{\bf x_1}
[\frac{q^4}{r^{2}}{\rm exp}(-2k_{D0}r)(1-B\frac{k_{D0}r}{n_{0}}\cos[\frac{\mathbf{p}}{2}\cdot(\mathbf{x}_{1}+\mathbf{x}_{2})])]\cdot \frac{\partial f_{1}({\bf X_1})}{\partial {\bf
v_1}}f_{1}({\bf X_2})
\nonumber \\
\end{eqnarray}
where
\begin{equation}\phi_{12}^{2}= \frac{q^4}{r^{2}}{\rm exp}(-2k_{D0}r)(1-B\frac{k_{D0}r}{n_{0}}\cos[\frac{\mathbf{p}}{2}\cdot(\mathbf{x}_{1}+\mathbf{x}_{2})])\end{equation}
In the above expression we have neglected terms containing $(\frac{B}{n_{0}•})^{s}$ when $s\geq 2$.
The contribution of the first term of the $\phi_{12}^{2}$ has been calculated  in \cite{kn:ab3}. The value is given by
 \begin{eqnarray} &&\frac{n_{0}}{mk_B T}  \int  d{\bf X_2}\nabla_{\bf x_1}
[\frac{e^4}{r^{2}}{\rm exp}(-2k_{D0}r)]\cdot \frac{\partial f_{1}({\bf X_1})}{\partial {\bf
v_1}}f_{1}({\bf X_2})
\nonumber \\
&=&
\frac{e^{4}}{2mk_{B}Tk_{D}^{0}•}\frac{\partial }{\partial {\bf x_{1}}}[n(\textbf{x}_{1})]\cdot
\frac{\partial f_{1}({\bf X_1})}{\partial {\bf v_{1}}}
\nonumber \\ \end{eqnarray}

The calculation of the  contribution of the second part is 
\begin{eqnarray} &-&\frac{Bk_{D0}}{mk_B T}  \int  d{\bf X_2}\nabla_{\bf x_1}
[\frac{e^4}{r}{\rm exp}(-2k_{D0}r)(\cos[\frac{\mathbf{p}}{2}\cdot(\mathbf{x}_{1}+\mathbf{x}_{2})])]\cdot \frac{\partial f_{1}({\bf X_1})}{\partial {\bf
v_1}}f_{1}({\bf X_2})
\nonumber \\
&=&G\int d{\bf X_{2}}
\nabla_{\bf x_1}[\cos[\frac{\mathbf{p}}{2}\cdot(\mathbf{x}_{1}+\mathbf{x}_{2})] \int d{\bf k} \frac{e^{-i{\bf k}\cdot{(
{\bf{x_1-x_2}}) }}}{k^2+4 k_{D0}^2} ]\cdot \frac{\partial f_{1}({\bf
X_1})}{\partial {\bf v_1}}f_{1}({\bf X_2})
\nonumber \\
&=&G\int d{\bf X_{2}}
\nabla_{\bf x_1} [\cos[\frac{\mathbf{p}}{2}\cdot(\mathbf{x}_{1}+\mathbf{x}_{2})] \int d{\bf k} \frac{e^{-i{\bf k}\cdot{(
{\bf{x_1-x_2}}) }}}{k^2+4 k_{D0}^2} ] \cdot \frac{\partial f_{1}({\bf
X_1})}{\partial {\bf v_1}}\int d{\bf k_2} {\tilde f_{1}}({\bf
{k_2,v_2}})\exp(-i{\bf{k_2\cdot x_2}})
\nonumber \\
&=&\frac{G}{2}\int d{\bf v_{2}}
\nabla_{\bf x_1}  \int d{\bf k} \frac{e^{-i{\bf
k}\cdot{\bf{x_1}}}}{k^2+4 k_{D0}^2} \cdot \frac{\partial f_{1}({\bf
X_1})}{\partial {\bf v_1}}\int d{\bf k_2} {\tilde f_{1}}({\bf
{k_2,v_2}})[e^{i{\bf
\frac{p}{2}}\cdot{\bf{x_1}}}\delta {\bf (k+\frac{p}{2}-k_2)}+e^{-i{\bf
\frac{p}{2}}\cdot{\bf{x_1}}}\delta {\bf (k-\frac{p}{2}-k_2)}]
\nonumber \\\end{eqnarray}
where $G=-\frac{Bk_{D0}e^{4}}{mk_B T}$, $\tilde f_1({\bf
{k_2,v_2}}) $ is the Fourier transform of $f_1({\bf X_2})$. For
long wavelength perturbations such that $k \ll k_{D} $ , we can approximate
$$\frac{1}{k^{2} + 4k_{D0}^{2}} \simeq
\frac{1}{4k_{D0}^2}$$ 
\begin{eqnarray}
&=&\frac{G}{2}\int d{\bf v_{2}}
\nabla_{\bf x_1}  \int d{\bf k} \frac{e^{-i{\bf
k}\cdot{\bf{x_1}}}}{4 k_{D0}^{2}} \cdot \frac{\partial f_{1}({\bf
X_1})}{\partial {\bf v_1}} [e^{i{\bf
\frac{p}{2}}\cdot{\bf{x_1}}}{\tilde f_{1}}({\bf {k+\frac{p}{2},v_2}})+e^{-i{\bf
\frac{p}{2}}\cdot{\bf{x_1}}}{\tilde f_{1}}({\bf {k-\frac{p}{2},v_2}})]
\nonumber \\
&=&
-\frac{e^{4}}{mk_{B}Tk_{D0}}\frac{\partial }{\partial {\bf x_{1}}}[n(\textbf{x}_{1})[\frac{B}{4n_{0}}\cos(\mathbf{p}\cdot\mathbf{x}_{1})]\cdot
\frac{\partial f_{1}({\bf X_1})}{\partial {\bf v_{1}}}
\nonumber \\\end{eqnarray}

Finally, adding the contributions of eq.(11) and eq.(13)
\begin{equation}
n_{0}\int
 d{\bf X_2}{\bf a_{12}}\cdot \frac{\partial g_{12}}{\partial
{\bf v_{1}}}=\frac{e^{4}}{2mk_{B}Tk_{D0}}\frac{\partial }{\partial {\bf x_{1}}}[n(\textbf{x}_{1})[1-\frac{B}{2n_{0}}\cos(\mathbf{p}\cdot\mathbf{x}_{1})]\cdot
\frac{\partial f_{1}({\bf X_1})}{\partial {\bf v_{1}}}\label{v4}\end{equation}
We may define
\begin{equation}
\theta(\textbf{x}_{1})=\frac{1}{2n_{0}}[n(\textbf{x}_{1})[1-\frac{B}{2n_{0}}\cos(\mathbf{p}\cdot\mathbf{x}_{1})]\end{equation}

The kinetic equation containing effects of weak correlations is
given by the following :

\begin{equation}
{\bf v_{1}} \cdot \frac{\partial f_1({\bf X_1})}{\partial {\bf
x_{1}}}-\frac{\partial \phi({\bf
x_1})}{\partial {\bf x_1}}\cdot\frac {\partial f_1({\bf
X_1})}{\partial {\bf v_{1}}}+g\frac{\partial \theta({\bf
x_1})}{\partial {\bf x_1}}\cdot\frac{\partial f_{1}({\bf
X_1})}{\partial {\bf v_{1}}}=0\label{v5}\end{equation}

In  eq.(\ref{v5}), the variables $x,n,v,\phi$ are normalized by
$\lambda_{D0} (=n_{0}e^{2}/k_B T)$, $n_0$, $v_{th} (=\sqrt{k_B T/m})$, $k_B T$ respectively and
\begin{equation} g= \frac{1}{n_{0}\lambda_{D0}^{3}} \end{equation}

The equilibrium solution of eq.(\ref{v5}) is
\begin{equation}
f_{1}({\bf
X_1})=e^{-(v_{1}^{2}/2+\phi({\bf
x_1})-g\theta({\bf
x_1}))}\label{v6}\end{equation}

It is a Maxwell-Boltzmann distribution with an additional contribution of $-g\theta(\textbf{x}_{1})$ to the potential energy due to the correlation of the particles.

In an earlier attempt \cite{kn:ab3} we had obtained an expression of single particle distribution function in the context of weakly correlated inhomogeneos plasma system. Therefore, the difference between this work and \cite{kn:ab3} should be mentioned clearly. In \cite{kn:ab3} we had made an Ansatz that the correlation function which  depends only on the separation of the particles and the single particle distribution function in the homogenous plasma is also applicable to the Inhomogeneos systems. Therefore, in that case the inhomogeneity was observed in the single particle distribution function. The assumption is now lifted in this article as we have chosen the most general form of the correlation function obtained in \cite{kn:ab1} and the correlation function depends not only on the separation of the particles and the single particle distribution function  but also on the average position of the particles. Finally, it may be concluded that inclusion of correlation contributes to the potential energy of the system and the reults seems to be different atleast of the order of $B/n_{0}$ from the result obtained in \cite{kn:ab3}. At the end, it may be concluded that we have been able to incorporate the effect of inhomogeneity in a more proper way as it has been derived from a more accurate expression of the pair correlation function in the weakly inhomogeneous regime. This single particle distribution function may be utilized to study the thermodynamic properties of the concerned systems.

\end{document}